\documentstyle[12pt]{article}

\newcommand{\be}{\begin{equation}}
\newcommand{\ee}{\end{equation}}
\newcommand{\bea}{\begin{eqnarray}}
\newcommand{\eea}{\end{eqnarray}}
\newcommand{\ba}{\begin{array}}
\newcommand{\ea}{\end{array}}
\newcommand{\non}{\nonumber}
\newcommand{\PP}[2]{\; P^{(#1)}_{\, #2}}
\newcommand{\pp}[3]{\; p^{(#1,#2)}_{\, #3}}
\newcommand{\eee}[3]{\; e^{(#1,#2)}_{\, #3}}
\newcommand{\bb}[3]{\; b^{(#1,#2)}_{\, #3}}
\newcommand{\bbi}[3]{\; {b^{-\, }}^{(#1,#2)}_{\, #3}}
\newcommand{\bbpm}[3]{\; {b^{\pm\, }}^{(#1,#2)}_{\, #3}}
\newcommand{\id}{\mbox{$\; I\; $}}
\newcommand{\om}[1]{\omega^{(#1)}}
\newcommand{\sq}[1]{\sqrt{Q^{(#1)}}}

\newcommand{\BBpm}[1]{B^{\pm\,}_{\, #1}}
\newcommand{\BMA}{braid--monoid algebra}
\newcommand{\BWM}{Birman--Wenzl--Murakami}
\newcommand{\BWMA}{Birman--Wenzl--Murakami algebra}
\newcommand{\YBA}{Yang--Baxter algebra}
\newcommand{\YBE}{Yang--Baxter equation}
\newcommand{\YBO}{Yang--Baxter operator}
\newcommand{\TLA}{Temperley--Lieb algebra}

\begin{document}

\begin{center}
{\LARGE\bf Dilute \BWM\ Algebra \\[4mm]
           and D$^{(2)}_{n+1}$ models} \\[8mm]
{\large\sc Uwe Grimm} \\[2mm]
{\footnotesize Instituut voor Theoretische Fysica,
               Universiteit van Amsterdam,\\
               Valckenierstraat 65,
               1018 XE Amsterdam,
               The Netherlands} \\[6mm]
January 1994 \\[6mm]
University of Amsterdam Preprint ITFA--94--01 \\[8mm]
\end{center}
\begin{quote}
{\small\sf
 A ``dilute'' generalisation of the \BWMA\ is considered.
 It can be ``Baxterised'' to a solution of the \YBA .
 The $D^{(2)}_{n+1}$ vertex models are examples of corresponding
 solvable lattice models and can be regarded as the dilute version
 of the $B^{(1)}_{n}$ vertex models.
}
\end{quote}
\vspace*{2mm}

\section{Introduction}
\setcounter{equation}{0}

The theory of two-dimensional solvable lattice models
is intimately connected with a list of algebraic structures
with a wide range of applications in mathematics and physics
\cite{Jimbo89}.
Among those are e.g., the braid group \cite{YangGe89}
and the Temperley--Lieb
\cite{TempLieb} and Hecke algebras \cite{Martin91}.
The braid and Temperley--Lieb or monoid
\cite{Kauffman} operators were combined into a single
(so-called braid--monoid) algebra by
Birman and Wenzl \cite{BirWen} and independently
by Murakami \cite{Mura}, see also \cite{WadDegAku}.
Besides being related closely to solvable lattice models,
these algebras have another important property.
They admit a simple diagrammatic interpretation
in terms of transformations of strands or strings
and are of importance in the theory of knot and
link invariants (see e.g.\ \cite{WadDegAku}).
Recently \cite{GriPea93}, a generalisation
of \BMA s has been considered which
amounts to considering strings of different
``colours''. These are connected to
recently constructed critical solvable
lattice models \cite{WarNieSea92,Roche92,WarNie93,WarNieSea93}
which are related to (coloured) dense or dilute loop models.

A \BMA\ (also called knit or tangle algebra)
is defined as the algebra generated by
$b_{\, j}$, $b^{-1}_{\, j}$ and $e_{j}$
($1\leq j\leq N-1$, where $N$ corresponds to the number of
strings in the diagrammatic interpretation mentioned above)
subject to the following list of relations
\bea
b_{\, j} \: b^{-1}_{\, j} & = &
b^{-1}_{\, j}\: b_{\, j} \;\; = \;\; \id \non\\
b_{\, j}\: b_{\, k} & = & b_{\, k}\: b_{\, j}
\hspace{1cm} \mbox{for $|j-k|>1$} \label{braidrel} \\
b_{\, j}\: b_{\, j+1}\: b_{\, j} & = &
b_{\, j+1}\: b_{\, j}\: b_{\, j+1} \non\\[4mm]
e_{\, j}^{2} & = & \sqrt{Q}\: e_{\, j} \non\\
e_{\, j}\: e_{\, k} & = & e_{\, k}\: e_{\, j}
\hspace{1cm} \mbox{for $|j-k|>1$} \label{TLrel} \\
e_{\, j}\: e_{\, j\pm 1}\:e_{\, j} & = & e_{\, j} \non\\[4mm]
b_{\, j}\: e_{\, j} & = & e_{\, j}\: b_{\, j} \;\; = \;\;
\omega\: e_{\, j} \non \\
b_{\, j}\: e_{\, k} & = & e_{\, k}\: b_{\, j}
\hspace{1cm} \mbox{for $|j-k|>1$} \label{BMrel} \\
b_{\, j\pm 1}\: b_{\, j}\: e_{\, j\pm 1} & = &
e_{\, j}\: b_{\, j\pm 1}\: b_{\, j} \;\; = \;\; e_{\, j}\:e_{\, j\pm 1}
\non
\eea
where $\sqrt{Q}$ and $\omega$
are central elements (hence numbers in any representation)
and $\id$ denotes the identity.
(\ref{braidrel}) are the braid relations and
(\ref{TLrel}) the defining relations of the
\TLA\ \cite{TempLieb}.
Usually, two equations are added to these
relations, which are
\be
f(b_{\, j}) \;\; = \;\; 0 \;\; , \hspace*{1cm}
g(b_{\, j}) \;\; = \;\; e_{\, j} \;\; , \label{poly}
\ee
where $f$ and $g$ are some polynomials. The algebra
originally investigated in Refs.~\cite{BirWen,Mura}
(the \BWMA ) for instance
corresponds to the case where $f$ is a cubic and $g$ a quadratic
polynomial in the braids. The reason why these equations are listed
separately is that they do not have a diagrammatic representation
and that one might consider these as properties of certain
representations rather than as part of the defining relations
of the algebra.

Solvable lattice models are commonly constructed as
vertex or as
face or IRF (interaction-round-a-face) \cite{Baxter} models whose
Boltzmann weights satisfy the \YBE s.
This property can equivalently be stated in the form that
they define a representation of the
\YBA\ \cite{Baxter,AkuWad87b,DegWadAku88a} defined by
\renewcommand{\arraystretch}{1.2}
\be
\ba{l}
X_{\, j}(u)\: X_{\, j+1}(u+v)\: X_{\, j}(v) \;\; = \;\;
X_{\, j+1}(v)\: X_{\, j}(u+v)\: X_{\, j+1}(u)  \\
X_{\, j}(u)\: X_{\, k}(v) \;\; = \;\; X_{\, k}(v)\: X_{\, j}(u)
\hspace*{15mm} \mbox{for $|j-k|>1$} \ea
\label{YBAD}
\ee
\renewcommand{\arraystretch}{1}
where $X_{\, j}(u)$ ($u$ denotes the spectral parameter)
are local operators whose matrix elements
are the Boltzmann weights of the model, see
e.g.\ \cite{WadDegAku} for details. For vertex models, these
\YBO s (also called ``local face operators'')
are particularly simple as they act on an $N$-fold tensor
space ($N$ being the number of vertices in one row),
acting as the $R$ matrix (to be precise, as
$\check{R}(u)=PR(u)$)
at slots $j$ and $j+1$ and as the identity elsewhere.

Every crossing-symmetric (see e.g.\ \cite{WadDegAku}) representation
of the \YBA\ yields a representation of the \BMA\
by setting
\be
e_{\, j} \;\; = \;\;  X_{\, j}(\lambda)  \;\; , \hspace*{1cm}
b^{\pm\,}_{\, j} \;\; = \;\; k^{\pm 1}\: \lim_{u\rightarrow\mp i\infty}
\frac{X_{\, j}(u)}{\varrho(u)}\;\; ,
\ee
where $\lambda$ is the
crossing parameter and  $k$ and $\varrho(u)$ are appropriately
chosen normalisation factors.
Conversely,
one may be able to ``Baxterise'' \cite{Jones} a
representation of the \BMA\ to a representation of the
full \YBA . This is especially useful if one can find a general
expression for the \YBO\ in terms of the braids and monoids
which can be shown to fulfill the \YBA\ as a consequence of
the algebraic relations alone (maybe apart from
some additional assumptions,
for instance about the polynomial reduction relations (\ref{poly})).
In this way, every appropriate representation
of the \BMA\ gives rise to a solvable lattice model.

This paper is organised as follows.
First, we give a short
summary of the \BWM\ case which corresponds to a
\BMA\ where the braids satisfy a cubic reduction relation.
Representations of this algebra occur in the
$B^{(1)}_{n}$, $C^{(1)}_{n}$, $D^{(1)}_{n}$, and
$A^{(2)}_{n}$ series of vertex models \cite{Bazh85,Jimbo86}
and associated face models. This is of course well known
\cite{DegWadAku88b,CheGeLiuXue92}, but still
there are a few surprising observations which arise.
In Sec.~3, we introduce the dilute generalisation of the
\BWMA\ and a graphical interpretation in terms of
diagrams acting on strings of two kinds.
This algebra is then ``Baxterised'' \cite{Jones}
to a solution of the \YBA\ (\ref{YBAD}) in Sec.~4.
Here, the corresponding examples of known models
are the $D^{(2)}_{n+1}$ vertex models
\cite{Jimbo86}. The associated representation of the dilute
algebra can be regarded as a dilute version of the
\BWMA\ related to the $B^{(1)}_{n}$ models which becomes
more transparent by a suitable change of basis in the
expression of the $R$ matrix in Ref.~\cite{Jimbo86}.
Finally, the results are summarised in Sec.~5.

\section{\BWM\ Algebra}
\setcounter{equation}{0}

We assume that the braid satisfies the cubic
\be
(b_{\, j}-\sigma^{-1}\id)\:
(b_{\, j}+\sigma\id)\:
(b_{\, j}-\sigma\tau^2\id)
\;\; = \;\; 0
\label{cubic}
\ee
where the third eigenvalue is the twist $\omega$ and hence
\bea
\omega   & = & \sigma\tau^2 \non\\
\sqrt{Q} & = & 1 \; +\;\frac{\omega-\omega^{-1}}{\sigma-\sigma^{-1}} \\
e_{j}    & = & \id \; +\;
 \frac{(b_{\, j}-b^{-1}_{\, j})}{\sigma-\sigma^{-1}} \non
\eea
The \BMA\ (\ref{braidrel})--(\ref{BMrel})
with these additional relations
is known as the \BWM\ (BWM) algebra.
Then it is easy to show that
the following ansatz satisfies the \YBA\ (\ref{YBAD})
\cite{CheGeLiuXue92}
\be
X_{\, j}(u) \;\; = \;\; \id \; +\;
\zeta^{-1}\,\eta^{-1}\, (z-z^{-1})\,
(\tau^{-1} z b_{j} - \tau z^{-1} b_{j}^{-1})
\label{RBWM}
\ee
where $z=\exp(iu)$ ($u$ denotes the spectral parameter),
$\zeta=(\sigma-\sigma^{-1})$,
and $\eta=(\tau-\tau^{-1})$. It is crossing symmetric
with a crossing parameter $\lambda$ given by $\tau=\exp(i\lambda)$
(note that $X_{\, j}(\lambda)=e_{\, j}$),
and satisfies the inversion relation
\be
X_{\, j}(u)\: X_{\, j}(-u) \;\; = \;\;
\varrho(u)\,\varrho(-u)\,\id
\label{invrel}
\ee
with
\be
\varrho(u) \;\; = \;\; \zeta^{-1}\,\eta^{-1}\,
(\sigma z^{-1}-\sigma^{-1}z)\, (\tau z^{-1}-\tau^{-1} z) \;\; .
\label{rho}
\ee
Examples of solvable lattice models which
can be expressed in this form \cite{DegWadAku88b,CheGeLiuXue92}
are given by the
$B^{(1)}_{n}$, $C^{(1)}_{n}$, $D^{(1)}_{n}$, and
$A^{(2)}_{n}$ vertex models \cite{Bazh85,Jimbo86} and related
face models \cite{DegWadAku88b}.
In the notation of Ref.~\cite{Jimbo86}
(where the $R$ matrices are parametrised by $x=z^2$ and
a complex parameter $k$),
the corresponding values of $\sigma$ and $\tau$ are
\be
(\sigma ,\tau) \;\; = \;\; \left\{
\ba{l@{\hspace*{5mm}}l}
(k,\xi^{1/2}) & \mbox{for $B^{(1)}_{n}$ and $D^{(1)}_{n}$} \\
(-k^{-1},\xi^{1/2})  & \mbox{for $C^{(1)}_{n}$ and $A^{(2)}_{n}$}
\ea \right.
\ee
where as in Ref.~\cite{Jimbo86},
$\xi=k^{2n-1},k^{2n+2},k^{2n-2},-k^{n+1}$ for $B^{(1)}_{n}$,
$C^{(1)}_{n}$, $D^{(1)}_{n}$, and $A^{(2)}_{n}$, respectively.

An interesting observation is in order. None of the above expressions
in the \BWMA\
are altered by interchanging $\sigma\leftrightarrow-\sigma^{-1}$.
This means that for a given representation there are in fact {\em two}
\YBO s (which {\em a priori}\ \ are not the same
since the values of $\tau$ are different), the second one
given by Eq.~(\ref{RBWM}) with
\bea
\sigma   & \longrightarrow & \sigma^{\prime} = -\sigma^{-1} \non\\
\tau^{2} & \longrightarrow & (\tau^{\prime})^{2} =
-\sigma^{2}\tau^{2}\;\;.
\eea
Having a closer look at the examples provided by the vertex models
one realises that the pairs ($A^{(2)}_{2n}$,$B^{(1)}_{n}$) and
($A^{(2)}_{2n-1}$,$D^{(1)}_{n}$) are built on identical representations
of the \BWMA . This implies that face models
associated to $A^{(2)}_{2n}$ and $A^{(2)}_{2n-1}$
can be directly deduced from the corresponding
$B^{(1)}_{n}$ and $D^{(1)}_{n}$ models, respectively \cite{Kuniba91}.
Note that the $A^{(2)}_{2n-1}$ face models of Ref.~\cite{Kuniba91}
are actually built on $C^{(1)}_{n}$ and not
on $D^{(1)}_{n}$ (in contrast to the vertex models
of \cite{Bazh85,Jimbo86}),
compare the discussion in \cite{Kuniba91}
at the end of sec.~1.

Of course, one cannot obtain the recently constructed
dilute A--D--E models \cite{WarNieSea92,Roche92}
(which are related to
the $A^{(2)}_{2}$ (Izergin-Korepin \cite{IzKor}) $R$ matrix) in this way
since these have a different algebraic structure
(corresponding to a different gauge of the
$A^{(2)}_{2}$ $R$ matrix, see Ref.~\cite{GriPea93} for details)
as already mentioned above.

Surprisingly, there is no obvious partner for the
$C^{(1)}_{n}$ models. This either means that the second solution
defines a new additional series of solvable vertex models
(and corresponding face models) or that
they are related to other models
(for instance to fusion $R$ matrices
of one of the other series).
This question is left open here,
it certainly demands further clarification.

\section{Dilute Braid-Monoid Algebra}
\setcounter{equation}{0}

The idea of considering multi-colour generalisations
of \BMA s originates in the investigation
of recently constructed face models \cite{WarNieSea92,Roche92,WarNie93}
which are related to (coloured) loop models. In Ref.~\cite{GriPea93}
it was shown that these models could be conveniently
described in terms of two-colour generalisations of the
\TLA\ \cite{TempLieb}. This
has been the motivation to look for similar generalisations of
the \BWMA\ and associated solvable lattice models.

To generate the $m$-colour algebra, we need ``coloured''
braid and monoid operators $\bbpm{a}{b}{j}$,
$\eee{a}{b}{j}$ (where $a,b=1,2,\ldots,m$ denote
the colours) as well as projectors
$\PP{a}{j}$ which project onto colour $a$ at position $j$.
Also, $\sq{a}$ and $\om{a}$ become the colour-dependent
``Temperley--Lieb eigenvalue'' and twist.
Note that here we use superscripts ``$+$'' and ``$-$''
(of which the ``+'' is usually omitted)
to distiguish coloured braids and ``inverse'' braids,
see \cite{GriPea93} for details.
The full set of relations which defines the
algebra (for the general $m$-colour case) can
also be found in Ref.~\cite{GriPea93}, they
are straightforward generalisations of the one-colour
relations (\ref{braidrel})--(\ref{BMrel}).
In complete analogy to
the one-colour case, all the relations can be
interpreted graphically
where has to consider strings of two different colours
(see \cite{GriPea93}) which can never join.

Here, we are only interested in a
``dilute (two-colour) \BMA ''
by which we mean a two-colour
case where one colour (we choose colour ``2'')
is trivial in the sense
that
\be
\bb{2}{2}{j}\;\; =\;\;\eee{2}{2}{j}\;\; =\;\; \PP{2}{j}\PP{2}{j+1}
\ee
which implies $\sq{2}=1$ and $\om{2}=1$.
Furthermore,
\be
\bbi{a}{b}{j} \;\; = \;\; \bb{a}{b}{j}
\hspace*{5mm} \mbox{($a\neq b$).}
\label{b1b2}
\ee
This means that the only non-trivial
operators acting on two sites $j$ and $j+1$ are
$\bbpm{1}{1}{j}$, $\eee{1}{1}{j}$,
\mbox{$\pp{a}{b}{j}=\PP{a}{j}\PP{b}{j+1}$}
($a,b\in\{1,2\}$),
$\bb{a}{\tilde{a}}{j}$, and
$\eee{a}{\tilde{a}}{j}$ ($a\in\{1,2\}$,
$\tilde{a}=3-a$).

Thinking in terms of the graphical representation,
this means that the second colour can also be interpreted
as a vacancy of a string - but it is easier to draw
pictures with two types of strings as one has to keep
in mind where these vacancies are. Still, the special properties
of the second colour lead to a somewhat simplified graphical
representation than for the full two-colour algebra
(Eq.~(\ref{b1b2}) for instance means that one does not
have to distinguish between two types of crossings of
strings of different kind), see Ref.~\cite{Pea94}.

\section{Baxterisation of Dilute BWM Algebra}
\setcounter{equation}{0}

We now consider a dilute \BMA\ as introduced in the
preceding section where the subalgebra generated by objects of
colour ``1'' is of \BWM\ type.
Changing our notation of Sec.~2 slightly, we assume the
following cubic relation for the braids $\bb{1}{1}{j}$
\be
(\bb{1}{1}{j}-\sigma^{-1}\pp{1}{1}{j}) \:
(\bb{1}{1}{j}+\sigma\pp{1}{1}{j}) \:
(\bb{1}{1}{j}-\tau^2\pp{1}{1}{j}) \;\; = \;\; 0
\label{cubic1}
\ee
where the third eigenvalue is again the twist $\om{1}$
which yields
\bea
\om{1}   & = & \tau^2 \non\\
\sq{1}   & = & 1\; +\;\frac{\om{1}-(\om{1})^{-1}}{\sigma-\sigma^{-1}} \\
\eee{1}{1}{j} & = &  \id\; +\;\frac{(\bb{1}{1}{j}-\bbi{1}{1}{j})}
                              {\sigma-\sigma^{-1}} \non
\eea

The above relations together with the defining relations of the
algebra (see Sec.~3 and Ref.~\cite{GriPea93}) are sufficient to show
that the following ansatz satisfies the \YBA\  (\ref{YBAD})
\bea
X_{\, j}(u) & = &
 \; \pp{1}{1}{j} \non\\* & &
\mbox{} + \; \zeta^{-1}\,\eta^{-1}\,(z-z^{-1})\,
  \left(\tau^{-1}z\bb{1}{1}{j}-\tau z^{-1}\bbi{1}{1}{j}\right) \non\\* & &
\mbox{} + \; \eta^{-1}\, (\tau z^{-1}-\tau^{-1}z)\,
  (\pp{1}{2}{j}+\pp{2}{1}{j}) \non\\* & &
\mbox{} - \; \varepsilon_{1}\,
  \zeta^{-1}\,\eta^{-1}\, (z-z^{-1})\, (\tau z^{-1}-\tau^{-1}z)\,
  (\bb{1}{2}{j}+\bb{2}{1}{j}) \non\\* & &
\mbox{} + \; \varepsilon_{2}\,
  \eta^{-1}\, (z-z^{-1})\, (\eee{1}{2}{j}+\eee{2}{1}{j}) \non\\* & &
\mbox{} + \; \left(1\: -\: \zeta^{-1}\,\eta^{-1}\, (z-z^{-1})\,
           (\tau z^{-1}-\tau^{-1}z)\right)\, \pp{2}{2}{j}
\label{RDBWM1}
\eea
where as in Sec.~2, $z=\exp(iu)$,
$\zeta=(\sigma-\sigma^{-1})$,
$\eta=(\tau-\tau^{-1})$ and where
$\varepsilon_{1}^2=\varepsilon_{2}^2=1$ are two arbitrary signs.
The appearance of this freedom is actually trivial as
all relations of the dilute \BWMA\ are invariant under
the transformations
\mbox{$\bb{a}{b}{j}\rightarrow
      (-1)^{a-b}\bb{a}{b}{j}$} and
\mbox{$\eee{a}{b}{j}\rightarrow
      (-1)^{a-b}\eee{a}{b}{j}$}.

The expression (\ref{RDBWM1}) is manifestly crossing
symmetric with crossing parameter $\lambda$ defined by
$\tau=\exp(i\lambda)$. Note that in order to have the
crossing transformations of the braid and monoid operators as
suggested by the diagrammatic interpretation (see
Ref.~\cite{GriPea93}) one should use $\varepsilon_{2}=1$ in
Eq.~(\ref{RDBWM1}). This stems from the fact that the
mixed monoid operators $\eee{a}{b}{j}$
are crossing related to the mixed
projectors $\pp{a}{b}{j}$ ($a\neq b$)
which have a fixed sign due to the requirement that the sum of
the projectors gives the identity.
The inversion relation (\ref{invrel}) is satisfied by
(\ref{RDBWM1}) with
\be
\varrho(u) \;\; = \;\; \zeta^{-1}\,\eta^{-1}\,
(\sigma z^{-1}-\sigma^{-1}z)\, (\tau z^{-1}-\tau^{-1}z)
\ee
which looks exactly the same as Eq.~(\ref{rho}).

Comparing the above expression (\ref{RDBWM1})
with Eq.~(\ref{RBWM}) one observes
that not only the inversion relation but also
the part which only involves colour ``1'' has exactly the same
form as for the pure \BWM\ case. But
in both cases one has to keep in
mind that for a given representation,
$\tau$ (and hence $\eta$) has a different
meaning in the two expressions (\ref{RBWM}) and (\ref{RDBWM1}), because
the twist is given by $\om{1}=\tau^2$ here
whereas $\omega=\sigma\tau^2$ in the discussion of Sec.~2. Obviously,
the colour ``1'' part of Eq.~(\ref{RDBWM1}) alone does not
satisfy a \YBE .

Alternatively, Eq.~(\ref{RDBWM1}) (with \mbox{$\varepsilon_{1}=1$})
can be expressed in a more ``symmetric'' form which reads as follows
\bea
 X_{\, j}(u) & = &
 \eta^{-1}\, (\tau z^{-1}-\tau^{-1}z)\, \id  \non\\* & &
\mbox{} -\; \zeta^{-1}\,\eta^{-1}\, (z^{1/2}-z^{-1/2})\,
 (\tau z^{-1}-\tau^{-1}z)\, (z^{1/2}  B_{j} + z^{-1/2} B^{-1}_{j})
 \non\\* & &
 \mbox{} + \; \eta^{-1}\, (z^{1/2}-z^{-1/2})\,
 (\tau z^{-1/2}+\tau^{-1}z^{1/2})\, (\eee{1}{1}{j}+\eee{2}{2}{j})
\non\\* & &
\mbox{} +\; \varepsilon_{2}\,\eta^{-1}\, (z-z^{-1})\,
 (\eee{1}{2}{j}+\eee{2}{1}{j})
\label{RDBWM2}
\eea
Here, we used the same notation as in Eq.~(\ref{RDBWM1}) and
\bea
\id     & = & \pp{1}{1}{j} +\pp{1}{2}{j} +
              \pp{2}{1}{j} +\pp{2}{2}{j} \non\\*
\BBpm{j}  & = & \bbpm{1}{1}{j} +\bbpm{1}{2}{j} +
                \bbpm{2}{1}{j} +\bbpm{2}{2}{j}
\eea
We include this second form since it treats both colours
on a equal footing and might be more suitable for possible
generalisations.

As in Sec.~2, exchanging $\sigma\leftrightarrow -\sigma^{-1}$
leaves all algebraic expressions invariant.
But contrary to the former case
this does not lead to a different solution as the value of
$\tau$ (defined by $\om{1}=\tau^2$) is not affected by this
transformation and hence the \YBO\ is also unchanged.

The remainder of this section deals with the
$B^{(1)}_{n}$ and $D^{(2)}_{n+1}$ vertex models.
This follows a dual purpose:
on the one hand we want to show that the
$D^{(2)}_{n+1}$ models provide examples
for the algebraic structure defined above, on the other hand we will
show that the representations corresponding to the $D^{(2)}_{n+1}$
vertex models can easily be obtained from those related to
the $B^{(1)}_{n}$ vertex models. The reason why this is important is
simply that the same procedure should work for face models also
(at least in the trigonometric case).

Let us commence with the \BMA\ representation related to the
$B^{(1)}_{n}$ vertex models.
We define
\bea
b^{\pm\,}_{\, j} & = &
\id\otimes\id\otimes\ldots\otimes\id \otimes b^{\pm}\otimes
\id\otimes\ldots\otimes\id\otimes\id \non \\
e_{\, j} & = &
\id\otimes\id\otimes\ldots\otimes\id \otimes \: e\:\otimes
\id\otimes\ldots\otimes\id\otimes\id \label{bej}
\eea
where $b^{\pm}$ and $e$ act at positions
$j$ and $j+1$.
Using the notation of Ref.~\cite{Jimbo86},
the explicit form of the $d^{2}\times d^{2}$ ($d=2n+1$)
matrices $b^{\pm}$ and $e$ reads
\bea
b & = &  \;
\sum_{\alpha}\: k^{-1}\: \left(1\: +\: (k-1)\,
\delta_{\alpha,\alpha^{\prime}}\right)\:
E_{\alpha,\alpha}\otimes E_{\alpha,\alpha} \non\\* & &
\mbox{} +\; \sum_{\alpha\neq\beta}\: \left(1\: +\: (k-1)\,
\delta_{\alpha,\beta^{\prime}}\right) \:
E_{\alpha,\beta}\otimes E_{\beta,\alpha} \non\\* & &
\mbox{} -\; (k-k^{-1}) \:\sum_{\alpha<\beta}\:
E_{\alpha,\alpha}\otimes E_{\beta,\beta} \non\\* & &
\mbox{} +\; (k-k^{-1}) \:\sum_{\alpha>\beta}\:
k^{\bar{\alpha}-\bar{\beta}}\:
E_{\alpha^{\prime},\beta}\otimes E_{\alpha,\beta^{\prime}}
\label{repb} \\[4mm]
b^{-1} & = & \;
\sum_{\alpha}\: k\: \left(1\: +\: (k^{-1}-1)\,
\delta_{\alpha,\alpha^{\prime}}\right)\:
E_{\alpha,\alpha}\otimes E_{\alpha,\alpha} \non\\* & &
\mbox{} +\; \sum_{\alpha\neq\beta} \:\left(1\: +\: (k^{-1}-1)\,
\delta_{\alpha,\beta^{\prime}}\right)\:
E_{\alpha,\beta}\otimes E_{\beta,\alpha} \non\\* & &
\mbox{} +\; (k-k^{-1})\: \sum_{\alpha>\beta}\:
E_{\alpha,\alpha}\otimes E_{\beta,\beta} \non\\* & &
\mbox{} -\; (k-k^{-1})\: \sum_{\alpha<\beta}\:
k^{\bar{\alpha}-\bar{\beta}}\:
E_{\alpha^{\prime},\beta}\otimes E_{\alpha,\beta^{\prime}}
\label{repbi} \\[4mm]
e &  =  & k^{2n-1} \:
\sum_{\alpha,\beta}\: k^{\bar{\alpha}-\bar{\beta}}\:
E_{\alpha^{\prime},\beta}\otimes E_{\alpha,\beta^{\prime}}
\label{repe}
\eea
Here, $1\leq\alpha,\beta\leq d$,
$\alpha^{\prime}=d+1-\alpha$ ($d=2n+1$),
\be
\bar{\alpha} \;\; = \;\; \left\{
\ba{l@{\hspace*{5mm}}l}
\alpha + \frac{1}{2} & 1\leq\alpha\leq n \\
\alpha               & \alpha=n+1 \\
\alpha - \frac{1}{2} & n+2\leq\alpha\leq 2n+1
\ea \right.
\ee
and $E_{\alpha,\beta}$ are the $d\times d$ matrices
with elements
\mbox{$(E_{\alpha,\beta})_{i,j}=\delta_{i,\alpha}\delta_{j,\beta}$}.

These matrices fulfill the equations
\be
(b-k^{-1}\id)\, (b+k\id)\, (b-k^{2n}\id) \;\; = \;\; 0
\ee
and
\be
e \;\; = \;\; \id\; +\; \frac{b-b^{-1}}{k-k^{-1}}
\ee
and hence $b^{\pm\,}_{\, j}$
and $e_{\, j}$ (\ref{bej})
form a representation of the \BWMA\ with
\bea
\omega & = & k^{2n} \non\\
\sqrt{Q} & = & 1 \; +\; \frac{k^{2n}-k^{-2n}}{k-k^{-1}}
\eea
The corresponding \YBO\ (\ref{RBWM}) with
$\sigma=k$ and $\tau=k^{n-1/2}$ yields exactly the $R$ matrix
of Ref.~\cite{Jimbo86} (with $x=z^2$).

In order to obtain a representation of the dilute
\BWMA , we add one extra state to the local spaces
which is going to correspond to the second colour. The
corresponding matrices which act on the tensor product
of two spaces now have
dimension $(d+1)^2\times(d+1)^2$.
The (two-site) projectors $p^{(a,b)}$ are given by
\mbox{$p^{(a,b)}=P^{(a)}\otimes P^{(b)}$} with
\bea
P^{(1)} & = &
\sum_{\alpha} \: E_{\alpha,\alpha} \non\\*
P^{(2)} & = & E_{d+1,d+1}
\eea
where here and in what follows the summation variables
are restricted to values $1\leq\alpha,\beta\leq d$
which correspond to the states of colour ``1'' and
the now $(d+1)\times(d+1)$
matrices $E_{\alpha,\beta}$ are defined as above.
The representation matrices for the
\BWM\ part (which is the part that involves colour ``1''
only) ${b^{\pm\,}}^{(1,1)}$ and $e^{(1,1)}$
are just given by the same expressions as
the matrices $b^{\pm}$ (Eqs.~(\ref{repb}) and (\ref{repbi}))
and $e$ (\ref{repe}), respectively, but they now are of
course $(d+1)^2\times(d+1)^2$ matrices as well.
The mixed braids $b^{(1,2)}$ and $b^{(2,1)}$ are
\bea
b^{(1,2)} & = & \sum_{\alpha}\:
E_{d+1,\alpha}\otimes E_{\alpha,d+1} \non\\*
b^{(2,1)} & = & \sum_{\alpha}\:
E_{\alpha,d+1}\otimes E_{d+1,\alpha}
\eea
and the mixed Temperley--Lieb operators
have the following form
\bea
e^{(1,2)} & = & -k^{n+1}\: \sum_{\alpha}\: k^{-\bar{\alpha}}\:
E_{d+1,\alpha}\otimes E_{d+1,\alpha^{\prime}} \non\\*
e^{(2,1)} & = & -k^{-(n+1)}\: \sum_{\alpha}\: k^{\bar{\alpha}}\:
E_{\alpha^{\prime},d+1}\otimes E_{\alpha,d+1}
\eea
with $\alpha^{\prime}$ and $\bar{\alpha}$ defined as before.

The above equations define a representation of the
dilute \BWMA\ with
\bea
\om{1} & = & k^{2n} \non\\
\sq{1} & = & 1\; +\; \frac{k^{2n}-k^{-2n}}{k-k^{-1}} \non\\
\om{2} & = & \sq{2} \;\; = \;\; 1
\eea
Correspondingly, we obtain a representation of the
\YBA\ from Eqs.~(\ref{RDBWM1}) or (\ref{RDBWM2})
with $\sigma=k$ and $\tau=k^{n}$ and hence a
solvable vertex model with $2n+2$ states.

As it turns out, these models are just the
$D^{(2)}_{n+1}$ vertex models
which play a somewhat singular role in Ref.~\cite{Jimbo86}
as they are the only series of $R$ matrices which
do not commute at one place (i.e., in general
\mbox{$[\check{R}(u),\check{R}(v)]\neq 0$}) which
already implies that the \YBO\ cannot be written as
a polynomial in a braid operator alone.
The expression obtained
here (choosing \mbox{$\varepsilon_{1}=\varepsilon_{2}=1$}
in Eqs.~(\ref{RDBWM1}) or (\ref{RDBWM2}))
is related to $\check{R}(x)=P R(x)$ of Ref.~\cite{Jimbo86}
(with $x=z$, $\xi=k^{n}=\tau$, $P$ is the permutation map
$P: v\otimes w\mapsto w\otimes v$) by an
orthogonal transformation with the  matrix
$S\otimes S$ where $S=S_{1}\cdot S_{2}$ with
\bea
S_{1} & = &   \sum_{\alpha=1}^{n+1}\: E_{\alpha,\alpha} \;\; +\;\;
              \sum_{\alpha=n+2}^{d}\: E_{\alpha,\alpha+1} \;\; +\;\;
               E_{d+1,n+2} \non\\
S_{2} & = &   \sum_{\alpha=1}^{n}\:  E_{\alpha,\alpha}\;\; +\;\;
              \sum_{\alpha=n+3}^{d+1}\:  E_{\alpha,\alpha} \non\\*
& & \mbox{} +\;\;
              \frac{1}{\sqrt{2}}\: \left(
              E_{n+1,n+1} +
              E_{n+1,n+2} +
              E_{n+2,n+1} -
              E_{n+2,n+2} \right)
\eea
i.e., the additional colour ``2'' state
corresponds in Jimbo's basis \cite{Jimbo86}
to the asymmetric combination of states $n+1$ and $n+2$.
In particular, the projectors onto the two colours are not
diagonal in that basis.

\section{Summary and Outlook}
\setcounter{equation}{0}

A ``dilute \BWMA '' has been defined as a generalisation of the
well-known \BWMA\ \cite{BirWen,Mura}.
This was done following the general ideas of
Ref.~\cite{GriPea93} on multi-colour \BMA s.
Similar to the \BWM\ case, the dilute algebra
can be Baxterised to a two-parameter solution of
the \YBA . This means that every appropriate matrix representation
of the dilute algebra defines a solvable lattice model.

As an example, the representation
of the \BWMA\ which corresponds to the $B^{(1)}_{n}$
vertex models  was considered explicitly and was
enlarged to a representation of the dilute algebra.
It turned out that the such obtained solvable vertex models
are the $D^{(2)}_{n+1}$ vertex models, the $R$ matrix
differing from that of Ref.~\cite{Jimbo86} just by a simple
similarity transformation.

There are a number of question raised by the results
of this paper.

The first, of course, is about the nature of the
``second'' series of solutions related to the
$C^{(1)}_{n}$ representations of the \BWMA\
(see the last paragraph of Sec.~2). It would come as
a surprise to the author if these would indeed
correspond to new solvable models. It has to be checked if they
do not in fact correspond to fusion $R$ matrices.

Another question concerns face models (IRF models \cite{Baxter})
related to the $D^{(2)}_{n+1}$ vertex models.
The result of Sec.~4 means that one can construct
such models (at least with trigonometric weights) on the
basis of the known $B^{(1)}_{n}$ models \cite{JimMiwOka88}
in a similar way as the dilute A--D--E models
\cite{WarNieSea92,Roche92,WarNie93,GriPea93} are related
to the ``usual'' A--D--E models.
To do this one has to find a dilute extension of the
corresponding representations of the \BMA .

Eqs.~(\ref{RDBWM1}) or (\ref{RDBWM2})  give a
solution of the \YBE\ for any representation of the
dilute \BWMA .
But if one has a representation of the \BWMA\ itself
it appears to be quite straightforward to generalise it to
the dilute case as one sees in the example of the
$D^{(2)}_{n+1}$ models in Sec.~4. These could be
constructed starting from the BWM representation provided
by the $B^{(1)}_{n}$ models.
On the other hand, we had three such series
in Sec.~2, the other two being related to the  $C^{(1)}_{n}$ and
$D^{(1)}_{n}$ models. It therefore seems plausible that
these should also give rise to corresponding series of dilute models
which on the first view
do not seem to fit in the list of known solvable models.
But even if it turns out that
they are related to known models (for instance
by a gauge transformation) these expressions might still be
of use. It is plausible that
as it happens in other cases (for instance for the
$A^{(2)}_{2}$ models
\cite{Kuniba91,WarNieSea92,Roche92},
see also \cite{WarNie93,WarPeaSeaNie})
there exist several series of solvable face models which are
related to the vertex model $R$ matrix in different gauges.

These subjects are currently been investigated.

\section*{Acknowledgements}
\setcounter{equation}{0}
This work was supported by an EC Fellowship
(Human Capital and Mobility Programme).
The author would like to thank
B.\ Nienhuis, P.\ A.\ Pearce and S.\ O.\ Warnaar
for helpful discussions.

\end{document}